\documentstyle[amsthm,amsmath,amsfonts,amssymb]{article}
\def\refer#1#2#3#4#5{#1:\ {\sl #2}\ {\bf #3}\ {(#4)}\ #5;\ }
\def\nl{\hfil\break}

\def\dti{\!\cdot\!}
\def\rref#1~{~(\ref{eq;#1})}
\def\dref#1~{~Definition~\ref{df;#1}}
\def\bs#1{$\boldsymbol{#1}$}
\def\mbs#1{\boldsymbol{#1}}

\def\mSs{{\cal S}_*}
\def\Ss{${\cal S}_*$}

\def\mbR{{\Bbb R}}

\def\mbI{{\Bbb I}}

\def\rQ{{\rm Q}}

\def\LHs{${\cal L(H)}_s$}
\def\mLHs{{\cal L(H)}_s}
\def\mH{{\cal H}}

\def\mLH{{\cal L(H)}}
\def\LH{${\cal L(H)}$}

\def\H{${\cal H}$}

\def\PH{$P({\cal H})$}
\def\mPH{P({\cal H})}

\def\cl#1{${\cal #1}$}
\def\mcl#1{{\cal #1}}

\def\bcD{${\boldsymbol{\mcl D}}_r$}

\def\d#1,#2~{$d_{#2}{#1}$}
\def\md#1,#2~{d_{#2}{#1}}
\def\D#1,#2~{$D_{#2}{#1}$}
\def\mD#1,#2~{D_{#2}{#1}}

\def\mN#1,#2~{\|#1\|_{#2}}
\def\N#1,#2~{$\|#1\|_{#2}$}

\def\mfk#1{{\frak #1}}

\def\fTs{$\mfk T_s$}
\def\mfTs{\mfk T_s}

\def\ome{$\omega$\ }
\def\mome{\omega}

\def\mphi{{\varphi}}

\def\mrh{\varrho}

\def\hh#1,#2,#3~{$\hat h_{\mfk#1}(#2,#3)$}
\def\mhh#1,#2,#3~{\hat h_{\mfk#1}(#2,#3)}

\def\ph#1,#2~{$\varphi_{#1}^{#2}$}
\def\mph#1,#2~{\varphi_{#1}^{#2}}
\def\cmrh#1~{{\varrho_{#1}}}
\def\crh#1~{$\varrho_{#1}$\ }

\def\pph#1,#2~{$\tilde\mph#1,#2~$}
\def\mpph#1,#2~{\tilde\mph#1,#2~}
\def\un#1,#2,#3~{${\rm u}_#1(#2,#3)$}
\def\mun#1,#2,#3~{{\rm u}_#1(#2,#3)}
\def\gQ#1,#2~{$g_\rQ(#1,#2)$}
\def\mgQ#1,#2~{g_\rQ(#1,#2)}

\def\mtQ#1,#2~{\tau^\rQ_{#1}#2}
\def\tQ#1,#2~{$\tau^\rQ_{#1}#2$}

\def\vv#1,#2~{${\mathbf v}_{#1}(#2)$}
\def\mvv#1,#2~{{\mathbf v}_{#1}(#2)}

\def\L#1~{$\pounds_{#1}$}
\def\mL#1~{\pounds_{#1}}

\def\Ca{$C^*$-algebra}

\def\wrt{with respect to\ }
\def\om#1,#2~{$\omega_{#1}^{#2}$}
\def\mom#1,#2~{\omega_{#1}^{#2}}

\def\noidt{\noindent}

\def\emm#1~{{\bf #1}\ind{#1}}
\def\glss#1~{#1\glo{#1}}

\newcommand{\bequ}{\begin{equation}}
\newcommand{\enqu}{\end{equation}}
\newcommand{\barr}{\begin{eqnarray}}
\newcommand{\earr}{\end{eqnarray}}

\newcommand{\glo}{\glossary}
\newcommand{\ind}{\index}
\newcommand{\rarw}{\rightarrow}

\swapnumbers
\theoremstyle{plain}
\newtheorem{lem*}[thm]{Lemma*}       
\theoremstyle{remark}

\begin{document} 
\title{ On Symmetries in Nonlinear Quantum Mechanics\thanks{
The paper is written for the Proceedings of  
 `Quantum Theory and Symmetries' (Goslar, 18-22 July 1999)
 (World Scientific, 2000),
 edited by H.-D. Doebner, V.K. Dobrev, J.-D. Hennig and W. Luecke}}

\author{ Pavel B\'ona\\ 
Department of Theoretical Physics, Comenius University \\ 
Mlynsk\'a dolina, 842 15 Bratislava \\ 
Slovak Republic} 

\maketitle

\begin{abstract}
It is shown how nonlinear versions of quantum mechanics can be refolmulated in 
terms of a (linear) \Ca ic theory. Then also their symmetries are described as
automorphisms of the correspondong \Ca. The requirement of
``conservation of transition probabilities'' is discussed. 
\end{abstract}

\section{A formulation of nonlinear quantum mechanics}

Nonlinear quantum mechanics (NLQM) is usually formulated in a form of
nonlinear Schr\"odinger equation (NLSchE), using concepts of the traditional
(linear) quantum mechanics (QM). Let \H\ be the Hilbert space of QM, and the
set of ``pure states'' is identified with the projective Hilbert space \PH,
i.e. the set of one--dimensional projections $P_\psi$\ on \H\ 
(\PH\ is identified also with the set of ``rays'' $\mbs\psi:= 
P_\psi\mH\ni\psi\neq0$,
i.e. to the set of one--dimensional complex subspaces $\mbs\psi
\subset\mH$), endowed with  the quotient  topology induced from \H. Then the 
considered class
of NLSchE can be formulated with a help of a selfadjoint operator valued
function $ \mbs\psi\ (\in\mPH)\mapsto \tilde H(\mbs\psi)\equiv\tilde
H(\mbs\psi)^*$\ defined on a dense subset $D(\tilde H)$ of \PH. 
Then the NLSchE is written as
\bequ\label{eq;1nlse}
i\dti\partial_t\psi_t = \tilde H(\mbs{\psi_t})\psi_t,\quad\psi_t\in D(\tilde
H(\mbs{\psi_t})).  
\end{equation}
We shall restrict our attention to such NLSchE\rref1nlse~ where the
``Hamiltonians'', i.e. the functions $\tilde H$\ are of a specific form. To
express it in a concise form, let us consider \PH\ as a submanifold of the
real Banach space of symmetric trace class operators \fTs. Then the
differential (resp. Fr\'echet derivative) $D_\mrh h$\ of a real--valued function
$h:\mfTs\mapsto\mbR$ (in points $\mrh\in\mfTs$, where it is well defined) is a
bounded real linear functional on the tangent space $T_\mrh\mPH$, if
$\mrh\in\mPH$\ (cf. the concept of
\bcD--generalized differential in~\cite{bon1}), and it corresponds to a 
symmetric linear operator.\footnote{This correspondence uses the duality
between \fTs\ and \LHs\ given 
by the trace of products of operators, and uses also a restriction to the
submanifold \PH; the present exposition is, however,
rather simplified.} We shall assume, in the sense of the mentioned
correspondence, that the operator--valued function $\mbs\psi\mapsto\tilde
H(\mbs\psi)$\ corresponds to a differential $D_{\mbs\psi}h$. This is the
case of several usual nonlinear modifications and/or approximations of QM,
cf.~\cite{bon1,ash&schil,bon2}. In such a case, the NLSchE\rref1nlse~ is a
form of classical Hamilton equations on \PH, where the Poisson brackets are
defined as the unique extension of ones given for functions of the form 
$h_A(P_\psi):=Tr(P_\psi A)$ by the relation
\bequ\label{eq;1pb}
\{h_A,h_B\}(P_\psi):=i\dti Tr(P_\psi[A,B])\equiv h_{i[A,B]}(P_\psi).
\end{equation}
 The specific case of Schr\"odinger equation (with the Hamiltonian $H=H^*$) of
the ordinary (linear) QM is obtained from the Hamiltonian function $h:=h_H$.

\section{Symmetries in linear QM} 

A general symmetry transformation $\Phi$ of QM is any bijection 
$\Phi:\mPH\rarw\mPH$\ conserving the ``transition probabilities'':
\bequ\label{eq;1trpr}
Tr\bigl(\Phi(P_\psi)\Phi(P_\mphi)\bigr)=Tr(P_\psi P_\mphi),\quad\forall
P_\psi,P_\mphi\in\mPH.
\end{equation}
For continuous one--parameter groups $t\mapsto\Phi_t$\ of such symmetries, the 
known Wigner theorem gives 
\bequ\label{eq;lin}
\Phi_t(P_\psi)\equiv e^{-itH}P_\psi e^{itH},
\end{equation}
for some selfadjoint $H$, determined up to an additive numerical constant
uniquely; this is, however, a flow corresponding to the linear theory with
Hamiltonian $H$. Hence, the relation\rref1trpr~ cannot be
satisfied in nonlinear NLQM.

It is possible, however, to reformulate\rref1trpr~ in a way that can be
extended to NLQM. It is done by considering one of the projections
in\rref1trpr~, say $P_\psi$, as representing a state $\mome_\psi$, but the
 second one will be considered as
an observable. The time evolution $A\mapsto\Phi_t^T(A)$\ of observables $A$\ 
is just the rewriting of the given Schr\"odinger
evolution $\Phi_t:\mome\mapsto\Phi_t\dti\mome$\ of states \ome\ into the 
``Heisenberg picture'', what is just the dual (i.e. transposed) transformation 
to that of states:
\bequ
Tr(P_\psi\Phi_t^T(A)):=Tr(\Phi_t(P_\psi)A)\equiv(\Phi_t\dti\mome_\psi)(A),
\quad\forall\ \{\psi,A,t\}.
\end{equation}
Then the condition\rref1trpr~ for $\Phi_t\mapsto\Phi$\ has the form:
\bequ\label{eq;2trpr}
\left(\Phi_t\dti\mome_\psi\right)\left(\Phi^T_{-t}(A)\right)=
Tr\left(\Phi_t(P_\psi)\Phi^T_{-t}(A)\right)=Tr(P_\psi A),
\end{equation}
what is trivially valid for any linear transformation $\Phi_t$, and its 
transposed  $\Phi^T_t$. As we shall see, a natural definition of ``Heisenberg
picture''--like transformations $\Phi_t^T$\ of observables in NLQM
corresponding to a given (nonlinear) flow $\Phi_t$\ on \PH\ makes\rref2trpr~  
{\em valid also for nonlinear flows} on \PH.
Clearly, such a ``trivialization'' of the relation\rref1trpr~ looses its
informative value:\rref1trpr~ implies possibility of unitary implementation
of $\Phi_t$, but\rref2trpr~ does not imply, perhaps, anything on the form of 
$\Phi_t$.

Let us note that the {\em transition probabilities interpretation
of}\rref1trpr~ can be traced back to the {\bs reduction postulate} of Dirac and
von Neumann for the process of measurement in QM: States $P_\psi$\ ``jump'' into
eigenstates $P_\mphi$\ of the measured observable corresponding to the obtained
numerical result (being equal to the corresponding eigenvalue) with 
probabilities
expressed by\rref1trpr~. If we accept these ``jumps'' as physical processes,
the usual interpretation of\rref1trpr~ is the natural requirement of 
invariance of
the probabilities of these processes \wrt any symmetry transformation, e.g.
\wrt the transition to another  reference frame. The
proposed interpretation connected with\rref2trpr~, on the other hand,
requires {\em invariance of all expectations} \wrt such simultaneous (symmetry)
transformations of states, and also of observables, that are mutually 
connected in the above described way.

\section{``Koopman--like'' linear reformulation of NLQM}

Let us remind first, how the classical Hamiltonian mechanics on a
(2n--dimensional) symplectic
manifold $(M;\Omega)$ can be (almost equivalently: up to ``measure zero
problems'') rewritten in a linear form according to~\cite{koopman}: Any 
symplectic flow
$\phi_t$\ on $M$ conserves the Liouville measure $\wedge^n\Omega$. If we
introduce the Hilbert space $\mH:=L^2(M,\wedge^n\Omega)$, as well as the
transformations
\bequ\label{eq;1koop}
\bigl(U^\phi_tf\bigr)(m):=f\bigl(\phi_t(m)\bigr),\quad f\in 
L^2(M,\wedge^n\Omega), 
\end{equation}
then $U^\phi_t$\ are
unitary and (under some continuity condition) are expressible by a linear 
selfadjoint
``Liouville operator'' $L_\phi:\quad U_t^\phi\equiv\exp(-itL_\phi)$. We have
obtained a linear dynamical system on infinite dimensional \H\ containing
(up to measure zero subsets of $M$) all the information about the given finite
dimensional (nonlinear) Hamiltonian system.

Let us now formulate, in a way ``similar'' to the Koopman's one, a linear
quantum theory (QT) containing a given NLQM as a subtheory. Here the \Ca ic
framework will be useful. We shall also generalize straightforwardly NLQM to a
dynamics of all density matrices.\nl\nl 
\bs{(a)}\ ``Quantum phase space'' consists of all density matrices
$\mrh\in\mSs:=\frak T_{+1}\subset\mfTs$. On \Ss, canonical Poisson
brackets are defined:
\bequ\label{eq;2pb}
\{f,h\}(\mrh):=i\dti Tr\bigl(\mrh[D_\mrh f,D_\mrh h]\bigr),\quad f,h\in
C^\infty(\mSs),
\end{equation} 
where the differentials $D_\mrh f,\dots$\ are considered as operators
according to the above mentioned identification:
$T^*_\mrh\mfTs\simeq\mLHs$.\vspace{8pt}\nl
\bs{(b)}\ One--parameter symmetry (evolution) groups
$\Phi_t:\mSs\rarw\mSs$\ are chosen to be the Hamiltonian flows
\wrt\rref2pb~. These evolutions contain also solutions of all the considered
NLSchE, if restricted to \PH.\vspace{8pt}\nl
\bs{(c)}\ The \Ca\ of observables is chosen to be $\mcl C:=C_b(\mSs,\mLH)
:=$the set of bounded--operator--valued functions continuous in some 
``convenient''
topologies on \Ss\ and \LH.\footnote{Let us note that a necessity of a 
state--dependence of observables is a consequence of nonlinearity of
transformations and of symmetry requirements of the 
type\rref2trpr~.}\vspace{8pt}\nl
\bs{(d)}\ Any Hamiltonian flow $\Phi$\ of (b) can be realized by a unitary 
cocycle $u^\Phi:\mbR\times\mSs\rarw \mcl{U(H)}$\ (:= the unitary group of
\H), $(t;\mrh)\mapsto u^\Phi(t,\mrh)$\ so that
\bequ\label{eq;1cocyc}
\Phi_t\dti\mrh:=\Phi_t(\mrh)\equiv u^\Phi(t,\mrh)\mrh u^\Phi(t,\mrh)^*.
\end{equation}
If the flow $\Phi$\ is determined by a Hamiltonian $h:\mSs\rarw \mbR$, then
the corresponding cocycle  $u^\Phi$\ can be chosen as the unique solution of 
NLSchE written in the form:
\bequ\label{eq;2nlse}
i\partial_t u^\Phi(t,\mrh)=D_{\mrh_t}h\dti u^\Phi(t,\mrh),\quad
u^\Phi(0,\mrh)\equiv\mbI,\quad\mrh_t:=\Phi_t(\mrh).
\end{equation}

\noidt\bs{(e)}\ The transformation $\Phi^T_t$\ of ``observables'' $\mfk f\in\mcl
C,\ \mfk f:\mrh\mapsto \mfk f(\mrh)\in\mLH$\ is then defined by a quantum
analogy of\rref1koop~:
\bequ\label{eq;1aut}
\bigl(\Phi^T_t\dti\mfk f\bigr)(\mrh):= u^\Phi(t,\mrh)^*\mfk
f\bigl(\Phi_t\dti\mrh\bigr)u^\Phi(t,\mrh),
\end{equation}
what defines a one--parameter (automorphism) group of (linear!)
transformations of the linear space \cl C\ of operator--valued functions on
the ``elementary state space'' \Ss.\footnote{Density matrices $\mrh\in\mSs$\
represent states of the quantum system called {\em elementary mixtures}.
They should be distiguished from {\em genuine mixtures} expressed by
probability measures on \Ss: The physical interpretations of different measures
with the same barycentre are mutually different in NLQM.}\vspace{8pt}\nl
\bs{(f)}\ Let us consider, e.g., only the states $\mome\in\mcl C_{+1}^*$\ on 
the \Ca\ \cl C\ of the form
\bequ\label{eq;state}
\mome(\mfk f)=\int_{\mSs}Tr\bigl(\mrh\mfk f(\mrh)\bigr)\ \mu_\mome(d\mrh),
\end{equation}
where $\mu_\mome$\ is any probability measure on \Ss.
The states corresponding to points $\mrh\in\mSs$\ are represented by the 
Dirac measures
$\mu:=\delta_\mrh$. Let us define the transformation
\[(\Phi_t\dti\mome)(\mfk f):=\left(\Phi^{TT}_t\dti\mome\right)(\mfk f)\equiv
\mome\bigl(\Phi^T_t\dti\mfk f\bigr),\]
i.e. the transposed map of the automorphism $\Phi^T_t$\ from\rref1aut~.
Then, according to\rref1aut~ and\rref1cocyc~, one obtains an extension
of\rref1cocyc~ to more general states:
\bequ\label{eq;2aut}
(\Phi_t\dti\mome)(\mfk f)\equiv\mome(\Phi^T_t\dti\mfk f)= \int
Tr\bigl(\Phi_t\dti\mrh\ \mfk f(\Phi_t\dti\mrh)\bigr)\mu_\mome(d\mrh).
\end{equation} 
Hence, also a nonlinear version of ``transition 
robability conservation'' is fulfilled: 
\bequ
 Tr\biggl(\Phi_t\dti\mrh\ \bigl(\Phi^T_{-t}\dti\mfk
f\bigr)(\Phi_t\dti\mrh)\biggr)\equiv
Tr\bigl(\mrh\mfk f(\mrh)\bigr). 
\end{equation}

\section{Summary}

Hamiltonian forms of NLQM are contained in a linear QT that is formulated in
terms of a \Ca\ \cl C\ of operator valued functions on the space of all density
matrices. Then the automorphism group of \cl C\ contains also nonlinear
symmetry groups of the considered quantum system. Although the usual
requirement of ``transition probabilities conservation'' leads to linear
transformations of the Hilbert space, a natural reinterpretation of this
requirement is extendable also to NLQM.

Let us note, that the theoretical scheme sketched above can be developed 
into a theory containing, as exact subtheories, besides the mentioned QM 
and NLQM, also CM, and
also various ``quasiclassical approximations'' (as are WKB, or time
dependent Hartree-Fock theory), cf.~\cite{bon1,ash&schil}; we call this theory  
EQM (:= extended quantum mechanics),~\cite{bon1}.


\section*{Acknowledgments} 
The author thanks for support to the organizers of this conference; he was
supported also from grant No. V2F20--G of Slovak grant agency VEGA.

\end{document}